\documentclass[manuscript]{copernicus}  

\nolinenumbers
\usepackage{color}
\usepackage[T1]{fontenc}
\usepackage{amsmath}

\begin{document}

\title{Condensate formation in collisionless plasma
}

{\author[1,3]{R. A. Treumann}
\author[2]{Wolfgang Baumjohann$^*$}
\affil[1]{International Space Science Institute, Bern, Switzerland}
\affil[2]{Space Research Institute, Austrian Academy of Sciences, Graz, Austria}
\affil[3]{Geophysics Department, Ludwig-Maximilians-University Munich, Germany\protect\\
$^*${(Correspondence to: Wolfgang.Baumjohann@oeaw.ac.at)}
}

}

\runningtitle{Condensate formation}

\runningauthor{R. A. Treumann \& Wolfgang Baumjohann}

\received{ }
\pubdiscuss{ } 
\revised{ }
\accepted{ }
\published{ }


\firstpage{1}

\maketitle

  

\noindent\textbf{Abstract}.-- 
{Particle condensates in general magnetic mirror geometries in high temperature plasma may be caused by a discrete resonance with thermal ion-acoustic background noise near mirror points. The resonance breaks the bounce symmetry, temporally locking the particles to the resonant wavelength. The relevant correlation lengths are the Debye length in parallel direction and the ion gyroradius in perpendicular direction.   } 

\section{Introduction}
The notion of condensate formation under high temperature collisionless plasma conditions has come into use recently in relation to evolving structures known as mirror modes \citep[cf., e.g.,][for their linear theory]{chandrasekhar1961,hasegawa1969,yoon2017}. There is a wealth on publications on mirror modes, observations and theoretical focussing on ion and electron modes, linear and nonlinear, including many different effects like finite-Larmor radii, electron anisotropy, dependence on composition and external conditions, coupling to other modes and importance in turbulence and particle acceleration, and evolution of mirror modes into chains. Listing all of them here is not the right place  \citep[a long though by no means exhaustive list can for instance be found in a recent paper][dealing with observation of chains of electron mirror modes]{yao2019}. Still the theory of mirror modes is not completely understood, in particular the role of electrons in their evolution. Generally the notion of mirror modes has been attributed to the ion mode, partly for historical reasons because of lack of high resolution instrumentation, partly because electrons have been believed to contribute just a minor modification. The self-consistent separate evolution of electron mirror modes as observed in \citep{yao2019,treumann2018} is therefore important as it shows that they indeed can evolve independently on the ion mode completely separate from it \citep{yao2019} or inside it \citep{treumann2018}, both times on typical electron scales. Their large amplitudes require conditions which go  beyond linear theory and are not covered by the nonlinear attempts hinted at above. Here we refer to our suggestion that electron dynamics in mirror geometries may be responsible for them by generating a condensate. This  applies to both ion and electron mirror modes separately. Whether it also applies to isolated magnetic holes which have occasionally been observed in the solar wind \citep[first in][]{turner1977,winterhalter1994,wang2020a,wang2020b,wang2020c} and at the magnetopause \citep[first in][]{luhr1987,treumann1990}, followed by many others is a different question. 

Condensates, if present, provide the basis for a semi-classical phase transition in dilute collisionless plasmas which explains the observed partial Meissner effect under the prevalent high temperatures \citep{treumann2019,treumann2021}. The question of how precisely such condensates form has, however, been left open so far. At low temperatures near the Fermi boundary condensate formation via Cooper-Schrieffer pairing of electrons, mediated by interaction with phonons, is at the heart of solid state physics \citep[cf., e.g.,][]{fetter1971} where it leads to metallic superconductivity \citep{bardeen1957}. Dilute plasmas at their high temperatures are already ideal conductors. In analogy to Cooper-Schrieffer pair formation  pairing might ignite condensation of particles. However, in high temperature plasmas this is probably unrealistic in view of the weak resonantly generated localized attracting electric fields but may be important in superdense quantum plasmas causing superconductivity there. Condensates in dilute classical plasma, if forming, indicate macroscopic correlations and organization. Re-treating the condensate problem here yields that under suitable magnetic mirror conditions realized in space plasmas discrete particle resonance with thermal ion sound noise suffices to generate condensates. Reference to pair formation is a possible secondary higher order effect though also weakly contributing to condensate formation. 

\subsection{Magnetic mirror geometry}
Magnetized high temperature plasmas are abundant in space and the universe in general. Normally there is little spectacular about them. However, when forming condensates they may evolve towards self-interaction, phase transition, and semi-classical macrostates.  

Consider the well-known bounce motion of charged particles in magnetic mirror geometry \citep[cf., e.g.,][]{roederer1970,kennel1966,paschmann2003} under conservation of their orbital magnetic moments $\mu=\mathcal{E}_\perp/B(s)$, where $\mathcal{E}=p^2/2m$ is  particle energy, $\mathbf{p}=m\mathbf{v}$ momentum, and $\mathbf{B}(s)$ the stationary magnetic field which converges with increasing parallel coordinate $s$ from $s_0=0$ towards $s=s_m$, the mirror point along the field \citep[cf., e.g.,][]{baumjohann1996}. We also assume the presence of a background spectrum of thermal noise \citep{rodriguez1975,lund1996,lucek2005}. This may be  broad band, but for being specific we in analogy to solid state physics assume that it consists of phonons which in plasma are ion-sound waves with dispersion $\omega(\mathbf{k})\approx kc_s/(1+k^2\lambda_D^2)^{1/2}$ \citep[cf., e.g.,][]{treumann1997}. Usually the wave number $k$ is much smaller than the inverse Debye length $\lambda_D^{-1}$, $k\lambda_D\ll1$ corresponding to long waves propagating at sound speed $c_s\approx \sqrt{T_e/m_i}$ with $T_e$ the electron temperature (in energy units) and $m_i$ ion mass. In addition to their cyclotron gyration the energetic particles  perform a bounce motion at frequency $\omega_b$ along the magnetic field with rapidly decreasing parallel speed until bouncing back from their mirror location $s_m$ according to
\begin{eqnarray}
v_\|(t)&=&v_{\|0}\cos \omega_bt, \qquad  v_{\|0}=\sqrt{2\mathcal{E}_0/m}\cos\alpha_0 \\
s(t)&=&s_m\sin\omega_bt, \qquad \omega_b^2=(2\pi/\tau_b)^2=\mu B_0''/m
\end{eqnarray}
with $\alpha_0$ the initial pitch-angle, $B(s)-B_0(0)\approx \frac{1}{2}B''_0(0)s^2$, and $'=\partial/\partial s$. The mirror point $s_m=v_{\|0}/\omega_b$ is reached at time $t=\tau_b/4$, which shows that the one-sided volume $s>0$ is filled with a continuum of mirror points $s_m(s,\alpha_0)$ whose locations along the field are a function of energy and the initial pitch angle distribution $f(\mathcal{E},\alpha_0)$ which we here leave unspecified.
Conservation of the magnetic moments implies that $v_\|(s\approx s_m)\ll v_\perp$ which close to  $s_m$ becomes about zero, and $\mathcal{E}_\perp(s\approx s_m)\approx \mathcal{E}_0$.  (As a side remark we note that in quantum physics such a bounce motion implies a harmonic bounce spectrum $\mathcal{E}_b=({\ell}+\frac{1}{2})\hbar\omega_b$ which abolishes the volume degeneracy of the Landau levels $\mathcal{E}_{Ln}=(n+\frac{1}{2})\hbar\omega_{ce}$ \citep{huang1973} with harmonic numbers $\ell=1,2,\dots<\omega_{ce}/\omega_{b}$, and cyclotron frequency $\omega_{ce}\gg\omega_b$.)

{One immediate consequence of the mirror process is that, given an initial velocity distribution in the equatorial plane of the magnetic mirror bottle, the magnetic minimum at location $s=0$ on any field line (magnetic flux tube), their will be a continuous distribution of mirror points $s_m (s)$ arising along the magnetic flux tube that starts in the equatorial plane. For instance, if the parallel particle velocities in the magnetic field minimum $s=0$ are subject to a Maxwellian distribution $f_M(v_{\|0})$, then it is easy to show by using the above expressions for the bounce motion that for given parallel velocity $\omega_bs_m=v_{\|0}=v_{0}\cos\alpha_0$ the distribution of mirror points with $s$ along any particular field line becomes trivially
\begin{equation}\label{fsm}
f_{\|M}[s_m(s)]\propto \exp\Big[-\frac{m\omega_b^2s_m^2}{2T_{e\|}}\Big(1-\frac{s^2}{s^2_m}\Big)\Big]
\end{equation}
It shows that the location of the mirror point is fixed by the Maxwellian, depending on $v_0$ and pitch angle $\alpha_0$ in the field minimum, and the distribution in velocity maps continuously to a distribution of mirror points along $s$. For the full distribution as function of initial energy this expression has to be integrated over the initial velocity distribution including the pitch angle distribution on $\alpha_0$ but not over space in order to construct the distribution of $s_m(s,\mathcal{E}_0)$. Mirror points thus fill the entire volume of the mirror bottle along each flux tube that crosses the equator with the given continuous distribution as function of location along a flux tube and initial energy distribution. Since each mirror point inhibits the mirroring electrons to propagate up to larger $s$, the electron distribution gradually dilutes towards increasing $s$ and magnetic field. This will have to be kept in mind when below discussing the generation of the condensate.} 

{Let $\mathcal{N}_{ml_m}$ be the number of electrons which arrive at mirror point $s_{ml_m}.$ This number is 
\begin{equation}
\mathcal{N}_{ml_m}=\mathcal{N}_0-\sum\limits_{l=1}^{l_m-1}\mathcal{N}_{ml}
\end{equation}
The sum (or integral) over all mirror points at smaller $s$ removes the mirroring electrons from contributing to the next mirror point. At every mirror point $s_{ml}$ along $s$ the fraction $\Delta_c$ of those electrons having their mirror points at $s_{ml}$, resonate and become locked to the wave, contributes to the condensate.  Their fraction $\Delta_c$  is an about constant property of the resonance. The number of electrons contributed by every field line to the condensate is thus given by
\begin{equation}
\mathcal{N}_{c}\approx{l_m}\Delta_c
\end{equation}
where $l_m$ is the number of mirror points along the field line in the magnetic mirror configuration, which presumably is a comparably large number. Let the length of the field line in the mirror configuration be $L_B$ and the parallel correlation length $\xi_\|\ll L_B$, then $l_m\approx L_B/\xi_\|\gg1$. Below we show that $\xi_\|\sim\lambda_D$ is of the order of the Debye length $\lambda_D$. Determination of $\Delta_c$ on the other hand requires precise knowledge of the locking/trapping process.}

\begin{figure}[t!]
\centerline{\includegraphics[width=0.75\textwidth,clip=]{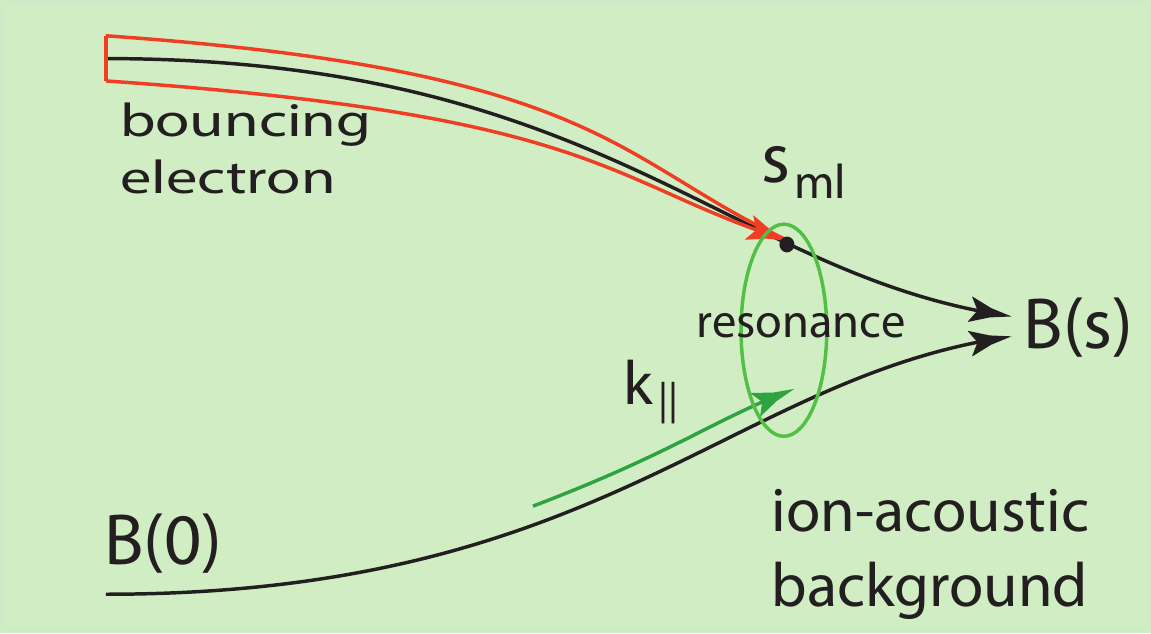}}
\caption{The underlying model: Schematic of mirror configuration, the $l$th mirror point $s_{ml}$, and resonance with an ion acoustic thermal background wave with parallel wavenumber $k_\|$. The green oval indicates the region of resonant correlation and coherence which around the particular mirror point is the Debye length in parallel direction and at least one ion gyroradius in perpendicular direction.} \label{fig-res1}
\end{figure}

\subsection{Discrete wave-particle interaction potential}
{The model underlying the following is sketched in Fig. \ref{fig-res1}.} The presence of a thermal ion sound background noise propagating at speed $c_s$ as, for instance, in the magnetosheath \citep[see][]{rodriguez1975,lucek2005} introduces an important subtlety. Propagation at long wavelengths $\lambda\gg\lambda_D$ is mainly along the magnetic field with $k\approx k_\|$. There is no purely perpendicular propagation, but propagation at larger oblique angles $k_\perp > k_\|$ is of slightly shorter wavelength \citep{treumann1997}. The ion-sound background enables the bouncing particles to fall into discrete particle resonance whenever their parallel bounce speed becomes comparable to the ion-sound speed $v_\|\approx c_s$ giving rise to an electric interaction potential \citep[cf., e.g.,][]{neufeld1955,krall1973} 
\begin{equation}\label{pot}
\Phi(\mathbf{x},t)= -\frac{e}{(2\pi)^3\epsilon_0}\int\frac{e^{i\mathbf{k\cdot}\mathbf{x}-i\omega t}\delta(\omega-{k_\| v_\|})}{k^2\epsilon(\mathbf{k},\omega)}\,d\mathbf{k}\,d\omega
\end{equation}
where $\mathbf{k}=(k_\|,\mathbf{k}_\perp)$ and $\mathbf{x}=(s,\mathbf{x}_\perp)$. For the mirror-trapped energetic bouncing particles this happens close to the mirror points $s\approx s_m$. The inverse of the response function $\epsilon(\omega_\mathbf{k},\mathbf{k})$ can for a large class of waves $\omega(\mathbf{k})=\omega_\mathbf{k}$ satisfying the dispersion relation $\epsilon(\omega_\mathbf{k},\mathbf{k})=0$ be written as 
\begin{equation}\label{eps}
\epsilon(\omega,\mathbf{k})^{-1}=\frac{k^2\lambda_D^2}{1+k^2\lambda_D^2}\Big(1+\frac{\omega_\mathbf{k}^2}{\omega^2-\omega_\mathbf{k}^2}\Big)
\end{equation}
For ion-sound waves in a magnetic field one has generally
\begin{equation}\label{disp}
\omega^2_\mathbf{k}=\frac{k_\|^2c_s^2\Lambda(\eta_i)}{\Lambda(\eta_e)+k^2\lambda_D^2}\Big(1+\frac{3T_i}{T_e}k^2_\|\lambda_{D}^2\Big), \qquad\Lambda(\eta)=I_0(\eta)e^{-\eta}
\end{equation}
where $I_0(\eta_{e,i})$ is the zero order modified Bessel function, and $\eta_{e,i}=\frac{1}{2}k_\perp^2r_{ce,i}^2$ with gyroradius $r_{ce,i}$ of electrons and ions. One usually assumes $k_\perp r_{ce}\ll1$ yielding $\Lambda(\eta_e)\approx 1$, which for ions is barely satisfied in our case but still will be assumed below. Also usually one assumes $T_i\ll T_e$ which simplifies the expressions.

In (\ref{eps}) the unity in the brackets accounts for Debye screening which here is of no interest. The test particle itself does not feel the screening which is important only for its environment, in particular, and outside the Debye sphere. The resonance in the second term plays a role only near the mirror point $s\approx s_m$ where the electron has very low speed $v_\|\approx c_s$ comparable to the ion sound velocity $\omega\approx k_\|v_\|$.  Here the particle energy is adiabatically transferred to gyration at cyclotron frequency $\omega_{ce}$. In the vicinity of all resonant particle mirror points the resonant term dominates the Debye screening (the unit in the brackets). The precise form of the potential felt by other particles of same sign is generally repulsive, but under restrictive conditions it can also become weakly attractive \citep{neufeld1955,nambu1985} at an approximate distance $\xi_D\approx 1.5\lambda_D$ \citep{treumann2019} in the wake of the moving particle outside the Debye sphere surrounding each of the bouncing particles. Clearly $\xi_D$ then plays the role of a ``correlation length'' providing the chance for transient electron pairing, the classical equivalent of Cooper-Schrieffer pair formation, which is interesting but of secondary importance in the non-solid  classical state where spin compensation becomes spurious. The number of paired particles always remains small and pairing is of minor importance in the dynamics of the plasma. 

\section{The condensate}
The main effect on the bouncing particles is the resonance near $s_m$ where the parallel particle energy becomes very small matching its resonance energy $\mathcal{E}_\mathit{res}\approx \frac{1}{2}mc_s^2$. The resonance temporarily violates the bounce invariance and locks the electrons to the ion-acoustic wave, mainly by trapping the electrons temporarily, thus for a while removing them from bouncing back into the equatorial weak field region:  
\begin{equation}\label{eqmo}
\frac{dv_\|(s)}{dt}\Big|_{s\approx s_m}=-\frac{e}{m}\frac{\partial}{\partial s}\Phi(s)\Big|_{s\approx s_m}-\omega_b^2(s-s_m),\qquad\omega_b^2=\frac{\mu B''_0}{m}
\end{equation}
Transformed to the mirror point location $s_m$ one has $v_\|(s_m)=c_s=$\,const. The mirror force  vanishes at $s=s_m$ while the particle is still carried ahead by the emerging resonant electric potential field that is induced by the ion sound.  In resonance the electron assumes the speed of the wave and rides together with the wave further up the field  being briefly fixed to the wave by the resonant potential. This could also be interpreted as trapping of the low energy bouncing particle in the ion-sound wave potential near the mirror point as discussed in another subsection further down. Here we first stick to the discrete wave particle resonance.

\begin{figure}[t!]
\centerline{\includegraphics[width=0.75\textwidth,clip=]{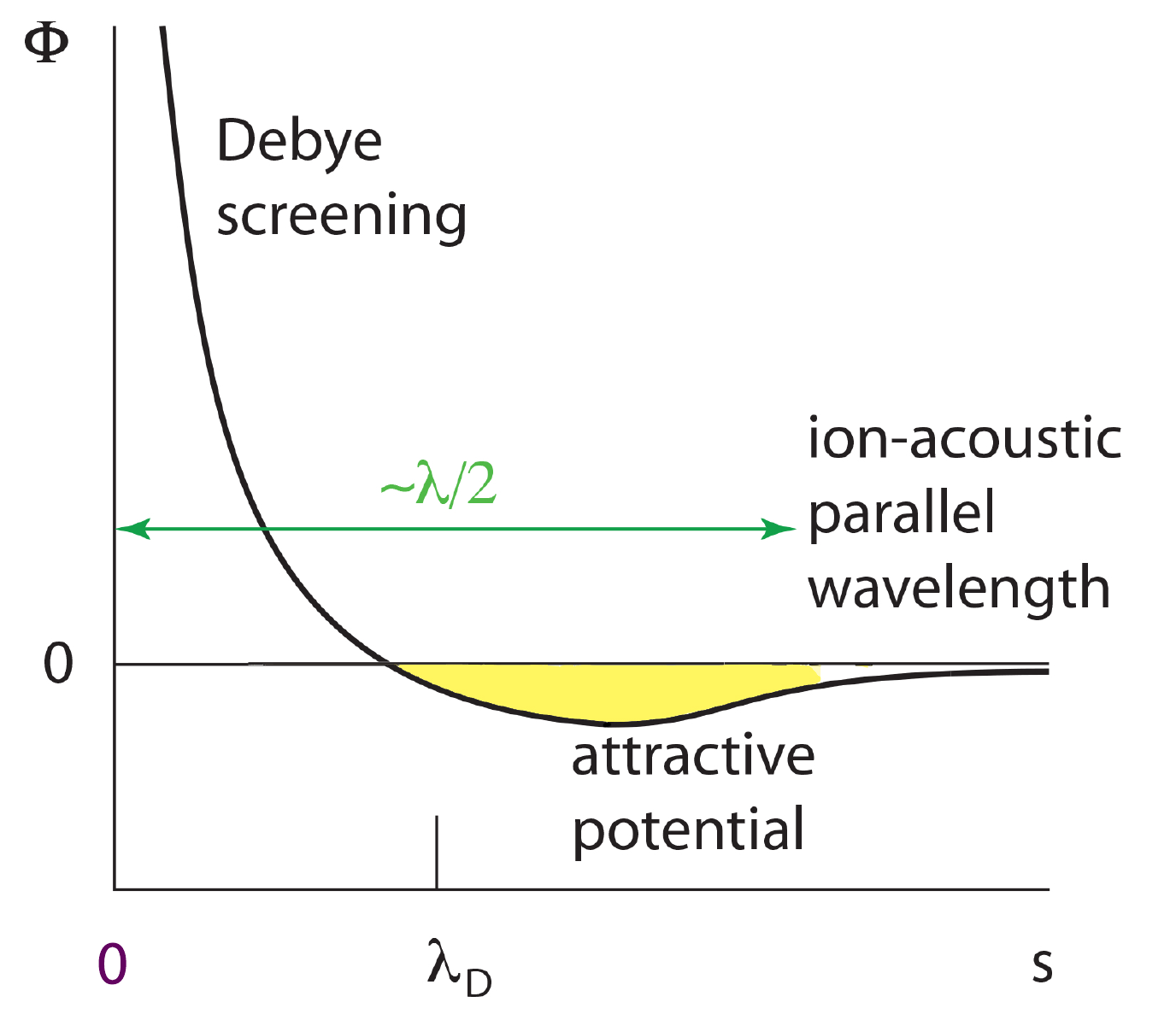}}
\caption{Debye and attractive potentials. The parallel ion-acoustic wavelength covers the attractive potential for resonant electrons. } \label{fig-att}
\end{figure}

\subsection{Interaction potential}
We  previously solved the above integral (\ref{pot}) for the potential $\Phi$ when  focussing on the general possibility of pair formation \citep{treumann2014}. Introducing the inverse response function it can be written
\begin{equation}\label{pot-a}
\Phi(\mathbf{x},t)=\frac{e\lambda_D^2}{(4\pi)^2\epsilon_0}\int\frac{\omega_\mathbf{k}\exp[ik_\|(s-v_\|t)]J_0(k_\perp\rho)}{1+k_\perp^2\lambda_D^2+k_\|^2\lambda_D^2}\Big[\frac{\delta(\omega-k_\|v_\|)}{\omega-\omega_\mathbf{k}}-\frac{\delta(\omega-k_\|v_\|)}{\omega+\omega_\mathbf{k}}\Big]dk_\|dk_\perp^2d\omega
\end{equation}
where the simplified ion-sound dispersion relation $\omega_\mathbf{k}\approx k_\|c_s$ has been used. The zero order Bessel function results from  $\theta$-integration over the interval $[0,2\pi]$ with respect to $k_\perp\rho\sin\theta$ in the exponential when adopting a Bessel representation. In the $\omega$-integration the two poles on the real $\omega$-axis must be surrounded depending on the convergence properties of the exponential. They are located on the real $\omega$ axis. For $s<v_\|t$ and $v_\|>0$ the integration path is deformed into the lower complex plane, adding a factor $i\pi$ to the integral. Otherwise for $s>v_\|t$ the deformation is into the upper half-plane and the factor becomes $-i\pi$. Integration with respect to $k_\|$ and combination of both cases gives
\begin{equation}
\Phi(\mathbf{x},t)=-\frac{e\lambda_D^2}{8\pi\epsilon_0v_\|}\int dk_\perp^2\frac{\omega_\mathbf{k}J_0(k_\perp\rho)}{1+k_\perp^2\lambda_D^2+\omega_\mathbf{k}^2\lambda_D^2/v_\|^2}\sin\Big(\frac{\omega_\mathbf{k}}{v_\|}\big|s-v_\|t\big|\Big)
\end{equation}
This potential is felt by the particle in interaction with the wave. One notices again that the electron does not feel its own screening. For itself the electron is naked under its clothes. Other particles outside the Debye sphere find the electron dressed.  

Figure \ref{fig-att} shows the transition from the Debye screening potential of the electron to the trapping potential (yellow negative potential part) at distance $\sim 1.5\lambda_D$ outside the screening Debye sphere along the parallel coordinate $s$. Electrons of sufficiently low relative speed outside the Debye sphere of the electron can become trapped in the negative potential trough. Though the potential is weak the trough contains a large number of electrons such that trapping and pair formation has a finite probability. Pair formation also depends on the direction of spin, but temporary trapping of electrons in the trough is always possible not forming pairs and thus occupying different energy levels.  

\subsection{Locking potential} 
The last form was useful in finding the interval where the potential would become attractive. Finally, integration with respect to the perpendicular wavenumber is possible with some further simplifications \citep{treumann2019} yielding
\begin{equation}
\Phi(s,\rho,t)\approx-\frac{e}{2\sqrt{2}\pi\epsilon_0}\frac{c_s}{v_\|}\frac{\exp(-\sqrt{2}\rho/\lambda_D)}{\lambda_D|{\sigma}|^2}\big\{\sin|\sigma|-|\sigma|\cos|\sigma|\big\}, \qquad{\sigma}=-(s-v_\|t)/\lambda_D
\end{equation}
{For small $\sigma<1$ this potential is negative $\Phi(|\sigma|)\propto -\frac{1}{2}|\sigma|$ at the start of the yellow domain in Fig. \ref{fig-att}.} The radius $\rho$ in the exponential is of the order of several times $\lambda_D$ because the potential induced by the electron does not extend very far beyond the Debye sphere in radial direction. The exponential is of order $0.01<\exp(-\sqrt{2}\rho/\lambda_D)<0.1$. To lowest order at $v_\|(s)\approx c_s$ close to $s\approx s_m$ the bounce motion is distorted and the electron becomes quickly locked to the wave. This happens at time $t\approx\pi/2\omega_b-\tau$ which gives $\omega_b\tau\approx c_s/v_{0\|}$ or for the mirror point distance $s_m-\delta s\approx c_s\tau\approx  c^2_s/\omega_b v_{0\|}$. At this location $\lambda_D|\sigma|\approx \delta(s-v_\|t)$ is small. Expanding the braces in the expression for the potential yields $\{\dots\}\approx |\sigma|^3/3$ which compensates for $|\sigma|^2$ in the denominator. The potential at resonance is obtained as
\begin{equation}
\Phi(\delta s,\rho,\tau)\approx-\frac{e\,|\delta s-c_s\tau|}{6\sqrt{2}\pi\epsilon_0\lambda_D^2}e^{-\sqrt{2}\rho/\lambda_D} \to-\frac{e\,\exp{(-\sqrt{2}\rho/\lambda_D)}}{6\pi\sqrt{2}\lambda_D\epsilon_0}\sqrt{\frac{m}{m_i}}\frac{\omega_e}{\omega_b}\frac{c_s}{v_{0\|}}
\end{equation}
This potential is switched on when resonance sets on in the vicinity of the mirror point, acting at least for a time $\omega_b\tau$. It is clearly rather weak. However, being fixed to the wave it moves with it up the field such that it is felt by the electron whose parallel kinetic energy at the mirror point becomes the resonant energy. The potential comes up for the remaining  velocity difference $\delta v_\|\approx c_s-v_\|(\tau_b/4-\delta\tau)$ and enables locking the electron. The velocity difference $\delta(c_s-\delta v_\|)|_{s_m}$ is of second order in $\omega_b\delta\tau$, where $\delta\tau$ is the variation of $\tau$ caused by the velocity difference. An estimate gives
\begin{equation}
(\omega_b\delta\tau)^2\sim \frac{c_s^2}{v_{0\|}^2}\approx \frac{m}{m_i}\frac{T_e}{\mathcal{E}_0\cos^2\alpha_0}\lesssim10^{-4}
\end{equation}
which yields a short time variation of roughly $\delta\tau\sim 0.05\tau_b$ which can be neglected with respect to $\tau_b/4$. 

\subsection{Locking distance}
The locking distance $\langle\Delta s\rangle=s_1-s_m$ beyond $s_m$ in the increasing converging magnetic field is determined by the growing mirror force which ultimately removes the electron from resonance at location $s=s_1$ and  parallel speed $v_\|(s_1)=0$.  The locking time $\langle\Delta\tau\rangle =t(s_1)-t(s_m)\approx \langle\Delta s\rangle/c_s$ and $\langle\Delta s\rangle=s_1-s_m$ can be estimated putting the left-hand side in (\ref{eqmo}) to zero
\begin{equation}
\langle\Delta s\rangle^2\approx\frac{c_s^2}{2\omega_b^2}, \qquad \Delta\tau\approx \frac{\sqrt{2}}{\omega_b}=\frac{\tau_b}{\pi\sqrt{2}}\approx0.2\tau_b
\end{equation}
The time the particle remains in resonance after passing its nominal mirror point is a fraction of the bounce period $\tau_b$. In this brief locked slow motion $v_{0\perp}\gg c_s$ the magnetic moment is conserved. The electron becomes perpendicularly heated $\langle\Delta\mathcal{E}_\perp\rangle\sim \mu B''_0\langle\Delta s\rangle^2/2\approx mc_s^2/2$, which is a small amount only, causing an increase in energy anisotropy $\mathcal{E}_\perp/\mathcal{E}_\|$ of just one unit only.  

\subsection{Locking potential ignoring pairing}
It is instructive to look for a different way of solving Eq. (\ref{pot}) by ignoring any pairing. Let us return to its already evaluated version (\ref{pot-a}). We sketch the somewhat lengthy and tedious though straightforward analytical calculation  here and only provide the final result. Straightforward integration of (\ref{pot-a}) with respect to frequency transforms the integral into one with respect to the parallel wave number $k_\|$. We insert the ion-sound dispersion relation (\ref{disp}) for $\omega_\mathbf{k}$ and neglect any magnetic effects but in the denominator of the dispersion relation allow for oblique propagation $\mathbf{k}=(k_\|,\mathbf{k}_\perp)$. One observes that the parallel wave number cancels in the unsplit quadratically singular term in (\ref{pot}) when inserting for the wave frequency as it is proportional to $k^2\lambda^2_D$. Defining $u\equiv c_s/v_\|<\sqrt{1+k_\perp^2\lambda_D^2}$ for electrons moving towards $s_m$ in the converging field allows to manipulate the singular denominator to separate $k_\|^2\lambda_D^2$. This leaves us with a squared singularity for the wave number $k_\|$ which can again be factorized yielding two separate linear poles in the complex $k_\|$ plane which turn out purely imaginary:  
\begin{equation}
k_\|\lambda_D|_\mathit{pole}=\pm i\sqrt{1+k_\perp^2\lambda_D^2-u^2}\equiv \pm i\Gamma
\end{equation}
Depending on the sign of the exponent $ik_\|(s-v_\|t)$ in the exponential of the integral in (\ref{pot-a}) either only the positive or negative pole contributes. For negative sign it is the positive exponent while the negative pole is to be neglected.  This produces a factor $-2\pi i$ in front of the integral. Using the imaginary pole in the denominator simplifies it to become just $u^2$, and only the  $k_\perp$-integral remains. To solve it one observes that $k_\perp\lambda_D<1$ which allows expanding the root in the argument of the exponential. The remaining integral is tabulated. Its value at resonance $u^2=1$ is
\begin{equation}
\Phi(\sigma,\rho)\approx \frac{e}{8\pi^2\epsilon_0\lambda_D}\frac{|\sigma|}{(|\sigma|^2+\rho^2/\lambda_D^2)^{3/2}}
\end{equation}
This potential is repulsive for any other electron. It applies to all scales inside the Debye sphere and outside with the exception of the above inferred narrow attractive region.The same expression holds at resonance for $\sigma>0$ where the opposite pole has to be used in the calculation. This simplified form ignores any attractive potential as the approximation made suppressed that weak effect. It shows however that a repulsive potential moves with the electron in resonance with the ion sound thus attracting the positive part of the ion sound wave if only the particle is in resonance. The potential decreases with increasing $|\sigma|$ and $\rho$ as predicted. It maximizes at $|\sigma|\approx\rho^2/3\lambda_D^2$. 

Multiplying with charge $e$ one obtains an estimate for the parallel velocity difference $\delta v_\|$ beyond $c_s$ which the potential can compensate
\begin{equation}
(\delta v_\|)^2 \approx \frac{e^2}{4\pi^2m\epsilon_0\lambda_D}\frac{|\sigma|}{(|\sigma|^2+\rho^2/\lambda_D^2)^{3/2}} \sim 10^2\Bigg\{{|\sigma|^{-2}\atop (\rho/\lambda_D)^{-3}} \qquad 
\end{equation}
For $\lambda_D\sim O(10\,\mathrm{m})$ the numerical factor is of the order of roughly $100$ m$^2/$s$^2$. Since both $|\sigma|>1$ and $\rho/\lambda_D>1$ the value depends on the ratio of the two quantities in the last expression. Hence the potential adjusts roughly for a difference in field-aligned speed of about $|v_\|-c_s|\sim10$\,m/s.

There is another point to make about the resonance. The thermal wave spectrum is usually isotropic which implies that waves parallel and antiparallel will be present in the plasma volume. Hence, once the mirror force takes over and re-accelerates the particle down the field, the particle if reaching its speed $-v_\|=-c_s$ can enter another resonance with a wave propagating anti-parallel to the field. Since the mirror force rapidly decreases during the particle displacement toward the magnetic minimum, the particle can stay for longer time in resonance until reaching the opposite mirror point $-s_m$ and becoming released  when having passed it. This is a substantial fraction of the trapped population in the condensate moving at constant speed $v_\|=c_s$ up and down the field. Effectively this process gradually causes parallel cooling and perpendicular heating of the condensate. The mean parallel temperature $T_{\|\mathit{cond}}$ of the condensate or its energy is then simply $\mathcal{E}_{\|\mathit{cond}}\approx mc_s^2/2\ll T_{e\perp}$. Condensate formation in this case becomes an efficient way of splitting the population into non-resonant isotropic and resonant  anisotropic parts, the latter forming the condensate.

\subsection{Electron trapping in resonance}
\begin{figure}[t!]
\centerline{\includegraphics[width=0.75\textwidth,clip=]{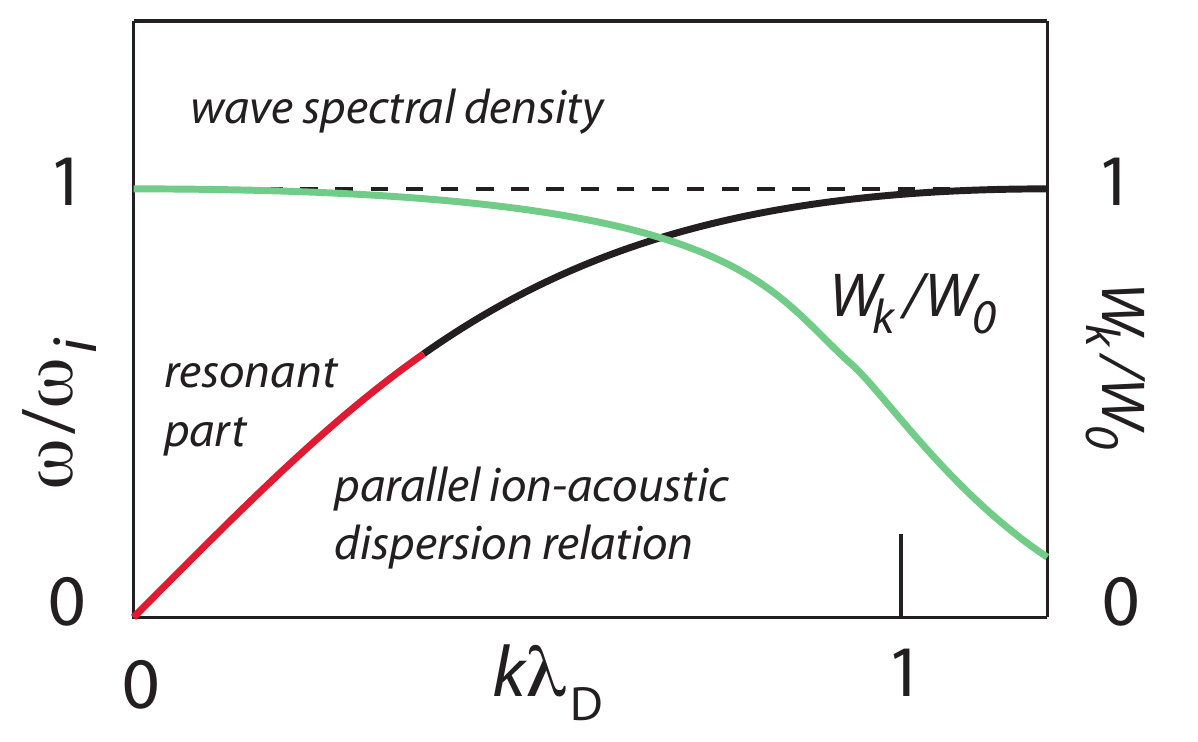}}
\caption{Parallel ion acoustic wave dispersion relation $\omega(k_\|)/\omega_i$ (with $\omega_i$ the ion plasma frequency) for fixed $k_\perp$, and ion acoustic thermal background noise spectrum $W_k/W_0$ as function of $k_\perp\lambda_D$. (Note that on the abscissa $k$ stands for both, $k_\|$ when referring to the dispersion relation, and $k_\perp$ when  meaning the spectral density.) The resonant part of the dispersion relation is shown in red. The spectral density (green) of the ion acoustic background noise is constant over the resonant range and decays like  $W_{k}\propto[1+k_\perp^2\lambda_D^2(1+k_\|^2/k_\perp^2)]^{-1}$ for fixed $k_\|$. For resonant electrons it is constant over several  parallel wavelengths and with $k_\perp\lambda_D\sim \lambda_D/r_{ci}<1$ up to a perpendicular distance of the ion gyroradius $r_{ci}$. } \label{fig-ia}
\end{figure}

Let us return to the interesting case of wave trapping of a single electron that has substantially slowed down near its mirror point $s_{ml}$. When the electron slows down to the ion-acoustic wave speed near $s_{ml}$ there is a good chance that the wave potential is large enough to trap the electron if only the wave can compensate for the residual particle energy. We recall the energy density of long wavelengths ion sound thermal fluctuations \citep{krall1973,baumjohann1996}
\begin{equation}
W_0\equiv\big\langle{\textstyle\frac{1}{2}}\epsilon_0|\delta E|^2\big\rangle\approx\frac{T_e}{\lambda_D^3}
\end{equation}
Figure \ref{fig-ia} shows the dependence of the ion-aoustic background noise spectrum on  the normalized wave number. Throughout the resonant part of the ion acoustic dispersion relation with phase velocity $\omega_k/k_\|\approx c_s$ (indicated in red) the wave power is practically constant being of value $W_0$. The energy difference near resonance $v_\|\approx c_s$ for a single electron in the parallel wavelength along the field is 
\begin{equation}
\frac{mc_s^2}{2}\Big(\frac{\delta v_\|}{c_s}\Big)^2\approx \frac{m}{m_i}T_e\Big(\frac{\delta v_\|}{c_s}\Big)^2
\end{equation}
Comparing the last two expressions in the long parallel resonant wavelength domain $k_\|\lambda_D<1$ we find that the originally fast electron can become trapped in the background noise at long wavelengths already at a location where the parallel speed difference of the electron to the ion sound is of the order $\delta v_\|< 10 c_s$. 

The parallel speed of the electron readily drops to a value which satisfies this rather weak condition which the long wavelength of the ion sound provides even at thermal wave levels. It is thus probable that an electron is trapped in the wave and stays in resonance as proposed above. A more precise calculation going beyond this order of magnitude estimate, which for the purposes of this communication suffices,  requires calculation of the probability of trapping in one of the thermal background ion acoustic waves which are distributed over a wide angular range of propagation with respect to the magnetic field, excluding perpendicular propagation. Since almost all electrons participate in bouncing, the fraction of trapping near mirror points depends predominantly on the distribution of the thermal ion acoustic wave spectrum along the magnetic field covering the continuous distribution of mirror points. It is thus reasonable to assume that the trapped particle fraction will be substantial providing a fairly dense condensate distributed over the entire fraction of the volume of mirror points inside the mirror geometry. 

{This case has an interesting consequence. The parallel ion acoustic wave length is somewhat larger than the Debye length $k_\|\lambda_D<1$, roughly amounting to $\lambda_\|\gtrsim 2\pi\lambda_D$ which for trapping of a slowed down electron in half a wavelength offers a good chance. This length covers a longer space interval than the pairing distance. It therefore includes  any pairs which have formed. The mutual  electron distance $N_0^{-\frac{1}{3}}$ is  much less than half the Debye length. In addition to pairs, a substantial fraction of electrons with $s_{ml}$ inside one half wavelength $\frac{1}{2}\lambda_\|\gtrsim\pi\lambda_D$ will therefore have the chance to become trapped, if only they meet the right wave phase. They all are mutually correlated, with parallel correlation length $\xi_\|\sim \lambda_D$. When becoming trapped they contribute to the condensate. One then has for the number
\begin{equation}
l_m= L_B/\xi_\|\sim L_B/\lambda_D
\end{equation}
which in the magnetosheath, for example, with typically at least $L_B\sim10^6$ m, and $\lambda_D\sim 10$ m, is of the order of $l_m\sim 10^5~\mathrm{to}~10^6$.} 

{Estimating the number of trapped electrons is  complicated. It should involve taking into account the spatial inhomogeneity of the distribution. Eq. (\ref{fsm}) suggests that, for a rather crude estimate, the fraction of electrons having their mirror points within one half wavelength centred at the mirror point $|s_m-s|\sim\xi_\|\sim\lambda_D$ is very roughly given by
\begin{equation}
\Delta_c\approx 1-\exp\Big(-\frac{\omega_b^2}{\omega_e^2}\frac{2s_{ml}}{\lambda_D}\Big) \approx\frac{\omega_b^2}{\omega_e^2}\frac{2s_{ml}}{\lambda_D}
\end{equation}
where use has been made of $\lambda_D=v_e/\omega_e$ (with thermal speed $v_e$ and plasma frequency $\omega_e$), and $s_{ml}+s\approx 2s_{ml}$ has been approximated. The frequency ratio $\omega_b/\omega_e\ll 1$ is a small number which, however, is compensated by the ratio of the mirror distance to the Debye length. In order to have an estimate let us assume $s_{ml}\approx L_B/2$  in the average, which is a reasonably conservative assumption. Each mirror point on the field line contributes a number 
\begin{equation}
\Delta_c\sim \frac{\omega_b^2}{\omega_e^2}\frac{L_B}{\lambda_D}
\end{equation}
of non-paired electrons to the condensate which, after summing up the contributions of all $l_m$ mirror points, yields an estimated average total contribution of the field line of
\begin{equation}
\mathcal{N}_c\approx \frac{\omega_b^2}{\omega_e^2}\frac{L_B^2}{\lambda_D^2}\approx 4\times10^4
\end{equation}
electrons to the condensate. As for an example magnetosheath conditions have again been assumed, with plasma frequency $\omega_e\sim 50$ kHz, electron cyclotron frequency about $\omega_c\sim 5$ kHz. The bounce frequency is a fraction of the latter, say $\omega_b\sim 10^2$ Hz only, thus $\omega_b/\omega_e\sim 2\times10^{-3}$. With $L_B\sim 10^6$ m and $\lambda_D\sim10$ m.}

{The above number of condensate electrons along the field line corresponds to a linear density of $N_{c\|}\sim 4\times10^{-2}$ m$^{-1}$. The linear density of the non-condensate background in the magnetosheath amounts to $N_{0\|}\approx 300$ m$^{-1}$. Hence the fraction of condensate to background electrons in the magnetosheath is roughly
\begin{equation}
N_c\approx 1.5\times 10^{-4} N_0 ~~\mathrm{m}^{-3}
\end{equation}
when assuming a magnetic mirror geometry containing an approximately homogeneous condensate. This fraction is small though not unreasonable. The importance of the condensate lies less in its density than in its property of being correlated and thus behaving coherently, which is at the base of Ginzburg-Landau (GL) theory \citep{ginzburg1950,bardeen1957,fetter1971,ketterson1999} of superconductivity. It is thus interesting that a similar effect is found in high temperature plasmas in the presence of a magnetic mirror symmetry. Locking of resonant electrons by electron trapping in ion-acoustic waves is thus a rather promising process for condensate formation and its consequences. In a more elaborate theory one should however in addition take into account the effect of the trapped electrons on the evolution of the ion-acoustic wave. Such processes have been widely considered for long time in nonlinear plasma theory \citep{sagdeev1969,davidson1972}. They could be applied directly here as they contribute to nonlinear growth of the wave which possibly reinforces electron trapping and maybe capable of amplifying the condensate formation and its macrophysical effect.}

\subsection{Correlation lengths}
The important property of the condensate is its internal correlation length $\xi$ which causes the condensate to behave approximately like a single particle occupying the same energy state, quantum mechanically spoken it has a unique common wave function $\psi$ which is made use of in Ginzburg-Landau (GL) theory and generally in the theory of superconductivity \citep{fetter1971} where it is called  the order parameter. It satisfies the first GL equation and its semi-classical version in application to mirror mode physics \citep{treumann2019}. As noted above, pairing  at $\xi_D\approx 1.5\lambda_D$ provides one such correlation length between the rather small, in fact tiny number of paired electrons even though by the argument of the last section a substantial fraction of electrons will have their mirror points in close contact with the wave and develop an attractive potential. Nevertheless the probability of pairing in the classical environment, where spin interactions play no role, will be small and thus by multiplication with the former probabilities reduce the importance of pairing. 

Of  greater importance is indeed the role played by the perpendicular ion sound wavelength. Propagating obliquely with respect to the magnetic field the perpendicular ion-acoustic wavelengths satisfy the condition $k_\|<k_\perp< r_{ci}^{-1}$ \citep{baumjohann1996,gary1993}. Wavelengths perpendicular to the magnetic field exceed the ion gyroradius. A substantial number of  bouncing electrons with magnetic mirror points $s_m$ located along this long perpendicular wavelength $\lambda_\perp$ are simultaneously in resonance $v_\|\approx\omega/k\approx c_s$ with the same ion-acoustic wave. They become trapped and locked to the wave by the resonance mechanism all together simultaneously moving at the same parallel speed $v_\|\approx c_s$ along with the wave parallel to the magnetic field. The entire set of these bouncing electrons is thus temporarily for their trapping time correlated not merely along the magnetic field but also in the perpendicular direction. The interparticle distance $N_0^{-1/3}\ll r_{ci}$ is very small compared with any of these correlation  lengths. Hence, in addition to the parallel coherence length $\xi_\|\sim\lambda_D$, the distance $k_\perp^{-1}\sim r_{ci}^{-1}$ acts as the relevant average perpendicular correlation length $\xi_\perp\sim k_\perp^{-1}\sim \langle r_{ci}\rangle$. All those particles behave coherently.  

\subsection{Ginzburg ratio}
In the further evolution of the condensate the important quantities are the magnetic penetration depth $\lambda_L= \sqrt{N_{0}/N_c}\lambda_i$, the London scale, and the correlation lengths $\xi$. Here $\lambda_i=c/\omega_i$ is the ion skin depth. Since the condensate density $N_c<N_0$ is  less than the ambient density, the London length is large. Inserting the number obtained above from particle trapping, we find that $\lambda_L\sim 10^{-2}\lambda_i$ which for the magnetosphere immediately suggests that mirror bubbles will only partially become depleted of the magnetic field. This is in excellent agreement with observation. Of utmost importance is, in addition, the so-called Ginzburg ratio $\kappa_G=\lambda_L/\xi_\perp$. It determines whether the macroscopic condensation effect is local or whether it affects the entire volume. 

The realistic case is of course the former where $\kappa_G>1$ which is realised in plasmas where the penetration depth of the magnetic field $\lambda_L\gg r_{ci}$ by far exceeds the ion gyro-radius, a situation realized in the magnetosheath and in any other observation of mirror modes. For the parallel correlation length, $\xi_\|=\lambda_D$ this is anyway trivially  given.

The identification of $\xi_\|=\lambda_D$ and in particular $\xi_\perp=r_{ci}$ as the relevant correlation lengths in condensate formation thus warrants that any effects the condensate will be responsible for modify the properties of the plasma locally only. For the identification of two such effects in magnetic mirror modes \citep[cf., e.g.,][]{lucek1999,{constantinescu2003},zhang2008,tsurutani2011,yao2019,yao2021} observed in the magnetosheath and solar wind, in particular the partial Meissner effect and its related phase transition, we refer to our previous work \citep{treumann2018}. The present note completes the theory by presenting the probable mechanism of condensate formation which in high temperature plasma to some extent is a surprise and possibly of farther reaching consequences showing that under particular purely classical conditions macroscopic effects similar to microscopic quantum states can arise and can be considered to resemble macroscopic quantum effects as they are subject to further affecting the dynamics. 

\section{Conclusions}
Single particle resonance near mirror points is a process which so far has been overlooked while possibly being capable of substantially changing the physics locally. Since in the always given presence of ion sound background noise in high temperature collisionless plasma it naturally will happen to a substantial fraction of the bouncing electrons, which should lead to the generation of a condensate as has been described in the previous sections. in magnetic mirror geometries condensate formation should therefore be a general effect if only a thermal wave background noise is present. This background noise can also be of different than of ion acoustic nature. For instance in the topside auroral ionosphere where in the geomagnetic mirror geometry particles bounce back and forth it should as well take place either with ion acoustic background noise or with any  Alfv\'en waves present whose phase velocity may match the parallel bouncing velocity of the energetic particles near their mirror points.  

The possibility for a correlated particle condensate under some peculiar conditions in high temperature collisionless plasma may be unusual. It will strongly become supported by trapping the slowed down parallel electrons in the thermal wave field as has briefly been discussed above. In all cases it  requires the presence of magnetic mirror geometries. Formation of a condensate has macroscopic consequences for the local plasma behaviour. In solid state quantum physics in analogy it lies at the basis of the microscopic BCS theory of  superconductivity \citep{bardeen1957}. Phenomenologically its dynamics  can in both cases be treated within GL theory which here, in the semi-classical approach, naturally applies \citep{treumann2019} when investigating the evolution of mirror modes in the magnetosheath.  

Condensate formation affects a non negligible fraction of the particle population which are correlated along the magnetic field over roughly few Debye lengths, and perpendicular to the magnetic field over one perpendicular wave length of the ion acoustic wave, corresponding to at least one ion gyroradius. In mirror modes the presence of the condensate causes the evolution of chains of mirror bubbles according to the intervention of the Ginzburg ratio $\kappa_G$. Interacting bubbles oscillate at Josephson frequency  \citep{josephson1962,josephson1964} which in the magnetosheath is expected to be in the infrared or sub-millimeter range \citep{treumann2021}, oscillation presumably there being too weak for detection however. 

Secondary effects like emission of observable radiation \citep[cf., e.g.,][]{treumann2020} in cyclotron harmonic bands \citep[see, e.g.,][for recent observations]{paschmann2003,malaspina2021}  are possible because of the large perpendicular anisotropy the condensate contributes . Application to reconnection, which evolves as well in the presence of a thermal background of ion-acoustic waves, will be investigated elsewhere, bearing in mind that under rather weak conditions the linear collisionless tearing mode should provide the required initial magnetic mirror geometry. This requires a rather weak tearing instability the conditions for which must be checked first. The possibility that a thermal background of tearing mode fluctuations might suffice to provide a background of mirror geometries can be abandoned \citep{kleva1982} because of the extremely low magnetic fluctuation level it can merely provide. 

In astrophysics magnetic mirror geometries are abundant in all kinds of magnetic stars, offering a wide field for condensates to develop and play a role, in particular in the generation of radiation. Aside from these, application to solid state physics is the most interesting, however. Immersing prospective superconducting materials into sufficiently strong magnetic mirror geometries at varying temperature might support condensate formation at increased temperature and thus cause higher temperature superconductivity.  This is an interesting problem which will be considered elsewhere. 

{Large amplitude mirror modes have long been measured magnetically as well as in pressure balance for long time (see the cited literature and the references therein). Also ion-sound waves are continuously present and have been observed regularly since their first detection \citep{rodriguez1975} where they already had been identified as thermal background noise.  Mirror points are distributed all over the magnetic mirror bottle. Hence the condensate is as well distributed. Its properties are, however, probably, very difficult to extract from the data. As they are only a fraction of the electron population, and this fraction has large anisotropy and all its energy in the perpendicular direction. Extraction from the total electron distribution is thus an intricate experimental problem. In fact this is the same problem as in metallic superconductivity where the condensate has never been ``seen'', just its Meissner effect.}

{Thus the important question arises concerning the experimental observability of the condensate. Condensate formation takes place in the vicinity of the mirror points $s_m$. These are homogeneously distributed over the mirror bottle, depending on the pitch angle distribution function in its centre, the field minimum. Thus condensate particles will be present in the whole mirror volume. They by themselves obey a high thermal anisotropy of the order of the ratio of ion to electron mass $\sim m_i/m$ resulting from the reduction of their parallel velocity to the velocity of the ion acoustic wave $v_\|\sim c_s$. This is their only signature as belonging to the condensate, because of the indistinguishability of electrons. Their fractional density $q=N_c/N_0<1$  may become substantial filling the whole mirror volume (with the exception of the equatorial region, i.e. the magnetic minimum as this lacks any mirror points and thus is least vulnerable to the partial Meissner effect).  However, experimentally it will be rather difficult to identify the condensate from measurements of the electron distribution as this can be done only via measuring the condensate anisotropy $A_{c}\approx m_i/m$. Though this is large, it presumably cannot be observed separately. The requirement of monitoring the parallel energy $\epsilon_\|$ in the very low energy regime and identifying a positive bump localized near $\epsilon\approx mc^2_s/2$ is certainly too strong! Hence, the condensate anisotropy will be hidden in the overall anisotropy.} 

{In fact, however, in \citep{treumann2018}, following the first unexplained detection of this effect in \citep{baumjohann1999}, we pointed on the almost persistent observation of the occurrence of a weak magnetic enhancement (or maximum) in the very centre of the (about symmetric) mirror bottle. It occurs once the spacecraft by chance crosses the centre of the bottle. According to the above discussion,  the quasi-superconducting Meissner effect should be weakend here, because of the lack of any mirror points, the high passing speeds of the bouncing electrons and the absence of any resonance with the wave background noise. One  expects that the field here is depleted least, which might be interpreted if not as the ultimate observational proof though as experimental support of our condensate formation theory.}

{That the global pressure anisotropy is little affected can easily be verified when looking into the contributions of the initial pressure anisotropy $A_0=P_{0\perp}/P_{0\|}-1$ and the condensate pressure anisotropy $A_c=P_{c\perp}/P_{c\|}-1$ to the total pressure anisotropy $A=A_0+A_c$. Here, in a very strict conservative treatment, the pressures are used in order to consider the effect of the fractions of electrons in the normal non-locked and condensate states. It suffices to consider the situation at some particular unspecified mirror point $s_m$. There the distribution consists of the normal passing (i.e. not mirroring) particles and the mirroring resonant condensate particles. It is then easy to show that the total anisotropy, accounting for the fractional contributions of the two constituents, is expressed as
\begin{equation}
A\approx A_0\Big[1+2q^2/(1-q)^2\Big]+2q^2/(1-q)^2 \to \frac{A_0}{1-q}
\end{equation}
The simplified form on the right is obtained when neglecting the second term and assuming a finite initial anisotropy $A_0=T_{0\perp}/T_{0\|}-1>0$. Thus, unless $q\lesssim 1$ the total anisotropy is not substantially effected. In case  $A_0=0$ initially, a weak anisotropy is produced  by the second term in this expression. For the above estimate of the magnetosheath value $q\sim10^{-4}$ it is  tiny and impossible to identify unless this estimate is unrealistic and will be corrected for $q$.  In any case, any observer would interpreted it as an original electron anisotropy $A_0$ from which it cannot be distinguished and therefore will be misunderstood and taken responsible for generating electron mirror modes, as has been done in the literature. This would however be a misconception because the electron mirror mode is small-scale the order of $\lambda_e$ and very low amplitude, as we have shown when using historical observations \citep{treumann2018} where the electron mirror mode becomes visible inside the ion mirror mode only. In most cases it will be drowned in the general level of magnetic fluctuations. It is improbable that the electron anisotropy would exhibit any sign of the condensate. As pointed out the role of the condensate is not in its anisotropy but in its coherence. By it it modifies London's penetration depth and the Ginzburg ratio with their macroscopic consequences which are independent on the anisotropy.}

{We have been strictly conservative when calling for pressure anisotropy. The situation of the anisotropy becomes substantially more favourable when the condensate electrons appear as a separate population independent of the non-condensate electrons with their large and pronounced anisotropy unaffected by the symmetric background and density ratio. Which stand has to be taken remains not known and undecided by now.} 

{Thus experimental verification by observation of the particle distribution is probably unrealistic. It requires precise measurement of the fluctuation in the parallel distribution function at $v_\|\sim c_s$ which seems unrealistic in view of the estimated condensate density. The only possibly promising spacecraft observations are  Magnetospheric Multi Scale (MMS)-data sets in the magnetosheath under conditions when mirror modes are observed. In this respect we point to very recent most interesting mirror mode observations \citep{yao2019,yao2021} in the magnetosheath and solar wind. Still, identification of the condensate will be an intricate problem. Other effects of the condensate are hardly imaginable. Its parallel energy is too low, its large separate anisotropy lies on the wrong side of the electron distribution for exciting any kinetic plasma waves. The only other effect to be expected is the modification of the ion acoustic background spectrum due to  resonance near the mirror point and electron trapping. This resonance implies interaction with the ion acoustic waves which so far has not yet seriously been investigated.}

\section*{Acknowledgments}
 This work was part of a brief Visiting Scientist Programme at the International Space Science Institute Bern. RT acknowledges the interest of the ISSI directorate as well as the generous hospitality of the ISSI staff, in particular the assistance of the librarians Andrea Fischer and Irmela Schweitzer, and the Systems Administrator Saliba F. Saliba. We acknowledge valuable discussions with M. Delva,  R. Nakamura, Z. V\"or\"os, M. Volwerk, D. Winterhalter, and T. Zhang.

\bibliographystyle{plain}

\end{document}